\input harvmac
\noblackbox

\def\bold#1{\setbox0=\hbox{$#1$}%
     \kern-.010em\copy0\kern-\wd0
     \kern.025em\copy0\kern-\wd0
     \kern-.020em\raise.0200em\box0 }

\lref\MP{R.S.~Millman and G.D.~Parker, {\sl Elements of Differential Geometry},
(Prentice-Hall, Englewood Cliffs, NJ, 1977).}
\lref\CH{
W.~Helfrich, Z. Naturforsch. {\bf 28C} (1973) 693;
P.~Canham, J. Theor. Biol. {\bf 26} (1970) 61.}
\lref\TGB{
S.R.~Renn and T.C.~Lubensky, Phys. Rev. A {\bf 38}, 2132 (1988);
{\bf 41}, 4392 (1990).}
\lref\scherk{
J.C.C.~Nitsche, {\sl Lectures on Minimal Surfaces}, trans. by J.M.
Feinberg, ed. by A.~Schmidt (Cambridge University Press, New York,
1989).}
\nfig\fone{Twist-grain-boundary structure for $+2$ screw dislocations.  The
horizontal lines are the smectic layers at the grain boundary, while the 
line at $\pm \alpha/2$ are the layers from the boundary.  The very dark
set of horizontal lines follows a single smectic layer across grain boundaries.
Note that at the grain boundary, the smectic layers are dilated by
$1/\cos(\alpha/2)$.} 
\Title{}{\vbox{\centerline{Minimal Surfaces, Screw Dislocations and}\vskip2pt
\centerline{Twist Grain Boundaries}}}

\centerline{
Randall D. Kamien\footnote{$^\dagger$}{email: 
kamien@physics.upenn.edu} 
and T.C.~Lubensky\footnote{$^\ddagger$}{email: tom@lubensky.physics.upenn.edu}}
\smallskip\centerline{\sl Department of Physics and
Astronomy, University of Pennsylvania, Philadelphia,
PA 19104, USA} 

\vskip .3in
Large twist-angle
grain boundaries in layered structures are often described by Scherk's first surface whereas
small twist-angle grain boundaries are usually described in terms of an array of
screw dislocations.
We show that
there is no essential distinction between these two descriptions and that,
in particular, their comparative energetics depends crucially on the core
structure of their screw-dislocation topological defects.  
\Date{27 August 1998; revised 2 November 1998}

Contrary to their name, minimal surfaces are widely observed.  It has been proposed \ref\ELT{S.P.~Gido, {\sl et al.}, Macromolecules, {\bf 26} 
(1993) 4506.}\ that lamellae in diblock copolymer systems form minimal
surfaces so as to minimize the interfacial area between blocks. 
Other systems, such as cubic phases \ref\cubic{D.W.~Bruce, {\sl et al.}, J. Phys. II France {\bf 5}
(1995) 289.}\ and
the smectic-Q phase \ref\LEV{A.-M.~Levelut, {\sl et al.}, J. Phys. II
France, {\bf 7} (1997) 981.}\ are also described in this way.
These minimal surfaces contain grain boundaries, which rotate the layer
normals.  
On the other hand, a twist-grain-boundary (TGB) in a smectic-A
liquid crystal phase is described in terms of a lattice of
screw dislocations \TGB .  We will show that these two different
descriptions lead to nearly identical geometrical surfaces even
though the philosophy behind their construction is quite different.

Minimal surfaces are simply surfaces of zero mean curvature.  Because
of this, they play an important role in the physics of complex fluids: 
their curvature energy always vanishes.  Though they may have periodically repeated
unit cells or, like Scherk's first surface \scherk , provide a continuous transition
between parallel planes at different angles, they do not have an inherent
length scale -- a uniform dilation of a minimal surface is {\sl still} minimal.  One
of the simplest minimal surfaces is the helicoid: a surface with height function
$h(x,y)= h_0 +[b/(2\pi)]\tan^{-1}(y/x)$.  This surface is smooth and continuous,
and there are {\sl no} constraints on the parameter $b$ giving the
pitch of the helicoid.  A screw dislocation in
a smectic-A, however, is a topological defect in a lamellar 
structure with a {\sl specified periodic spacing}
$d$.  It has a height function identical to that of a helicoid, though 
its pitch must be an integer multiple of the layer spacing.
As we discuss below, this constraint is most easily
implemented in an Eulerian picture in which the single-valued phase of the
mass-density wave is spatially dependent.  Moreover, a true dislocation has a singular core because
the layer spacing and thus the strain diverges there.  This costs energy
{\sl precisely} because there is a preferred layer spacing.  Thus the existence
of a well-defined periodic reference state distinguishes screw dislocations
from helicoids.  In this letter, we will explore the consequences of this distinction.

We start by considering Scherk's first surface. The
height function $h(x,y)$ is given implicitly via \scherk :
\eqn\escherk{x\sin\alpha = \ln\left[
{\cos\left[\cos\left({\alpha\over
2}\right)h(x,y) 
+ \sin\left({\alpha/2}\right)y\right]\over
\cos\left[\cos\left({\alpha\over
2}\right)h(x,y) 
- \sin\left({\alpha/2}\right)y\right]}\right]= 
\ln\left[{\cos\left({\bf N}_1\cdot
{\bf t}\right)\over\cos\left({\bf N}_2\cdot{\bf t}\right)}\right],}
where ${\bf t}\equiv\big(x,y,h(x,y)\big)$ and
${\bf N}_{1,2}=\big(0,\pm\sin(\alpha/2),\cos(\alpha/2)$ are unit
vectors.  
By using the rightmost expression in \escherk , it is easy to see that
as $x\rightarrow -\infty$ (for $0<\alpha<{\pi\over 2}$) the surface is the solution
of $\cos\left({\bf N}_1\cdot {\bf t}\right)=0$, 
{\sl i.e.}, Scherk's surface is a family
of parallel planes with layer normal ${\bf N}_1$ and spacing $d=\pi$.  Similarly,
as $x\rightarrow \infty$ Scherk's surface becomes a family of parallel layers with
normal ${\bf N}_2$ and spacing $d=\pi$.

Note that as $\alpha\rightarrow 0$, \escherk\ becomes
\eqn\escrew{x\sin\alpha \approx \ln\left[1 - 2y\sin\left({\alpha/2}\right)
\tan h\right] \approx -y\sin\alpha\tan h}
so that 
\eqn\ehh{h(x,y) = {\pi\over 2} + \tan^{-1}\left({y\over x}\right),}
which is a single helicoid with pitch $p=2\pi=2d$ located at the origin.  Since
planes at infinity are separated by $\pi$, this can be interpreted as a
$+2$ screw dislocation.  
Note, however, that the normal distance between the planes
diverges along the $\hat z$-axis, a fact that will prove to be
essential in our discussion of energetics. 
In biphasic materials, such as diblock copolymers,
these layers should be alternately identified as A-blocks and B-blocks.  This
would make the layer spacing at infinity $\tilde d=2d=2\pi$ and
thus the Burgers vector $b=2\pi$ would be a strength $+1$ dislocation \ELT .
However, in the following we will take each layer to be identical.  It will
only be convention as to the strength of the dislocation -- the actual
length of the Burgers vector is the appropriate invariant.

For finite $\alpha$, the intersection of the Scherk's surface with the plane $x=0$
consists of two sets of equally spaced lines: one parallel to the $\hat z$-axis
at positions $y_k=kl_d$ for integer $k$, with separation 
$l_d=\pi/\sin(\alpha/2) = d/\sin(\alpha/2)$
and a second set parallel to $\hat y$-axis at the heights
$h_k=kd'$ with separation $d'=\pi/\cos(\alpha/2)=d/\cos(\alpha/2)$.
Around the first set of lines, Scherk's surface is a local helicoid 
with height function
$h(x,y) = -[1/\cos(\alpha/2)]\tan^{-1}\left[
x\cos(\alpha/2)/\delta y\right]$
for small $\delta y=y-y_k$.  Thus these helicoids have pitch $p_z=2d'$ equal
to twice the separation between the planes parallel to $\hat y$ at $x=0$.  Similarly,
around the second set of lines there are helicoids: for small $\delta h=h-h_k$, $y(x,h)=
-[1/\sin(\alpha/2)]\tan^{-1}\left[x\sin(\alpha/2)/\delta h\right]$.  Thus
they have pitch $p_y=2l_d$ equal to twice the separation between planes parallel
to $\hat z$ at $x=0$.  Note that {\sl neither} $p_z$ or $p_y$ are integer
multiples of the layer spacing at infinity $d$.  The two sets of helicoids
can be interchanged through  
the ``dual'' transformation $h(x,y)\leftrightarrow y$
and $\alpha\rightarrow \pi-\alpha$ which leaves Scherk's surface invariant and
interchanges $d'$ and $l_d$.  
In the following we
shall show that if the (non-chiral) ground state of the system is lamellar, with
a preferred layer spacing, then energetics will break the geometric degeneracy
of the dual mapping by converting one set of helicoids into true
topological screw dislocations with cores. Finally, for comparison with
the smectic structures we will consider, we calculate the derivatives of $h(x,y)$:
\eqn\escherkdif{\matrix{{\partial_x h(x,y)}
&=  -\sin\left({\alpha/2}\right)
{\displaystyle \sin(2\theta)\over\displaystyle \cosh(2\gamma) - \cos(2\theta)}\cr
{\partial_y h(x,y)} &= 
\tan\left({\alpha/2}\right) {\displaystyle\sinh(2\gamma)\over\displaystyle \cosh(2\gamma)
-\cos(2\theta)}\cr}}
where $\theta\equiv y\sin\left(\alpha/2\right)$ and
$\gamma\equiv {1\over 2}x\sin\alpha$.  From these derivatives
and boundary conditions Scherk's surface can be reconstructed.

We will now consider a surface constructed via a linear superposition of screw 
dislocations (LSD) in a smectic-A phase, parallel to $\hat z$
with separation $l_d$ along the $\hat y$-axis.  
Recall that a smectic is described in terms of a complex scalar mass-density
wave
\eqn\ecomplex{\psi = \vert\psi\vert e^{iq\left[z-\bar u(x,y,z)\right]},}
where $q = 2\pi/d$ and $d$ is the layer spacing.  
While the surface $h(x,y)$ 
is a true height function expressed in {\sl Lagrangian} co\"ordinates, smectics
are described in terms of the phase variable $\bar u$, which is {\sl Eulerian}.  
To compare with
Scherk's surface we shall take $d=\pi$ here and in the following. 
Because $\psi$ must be single valued, the Burgers vector $b$ must be an integer
multiple of the layer spacing.  The (Lagrangian) surfaces of constant
phase $\bar h(x,y)$ are defined implicitly through
\eqn\econphase{\bar h(x,y)-\bar u\left(x,y,\bar h(x,y)\right) = kd,}
where $k$ is an integer.  
Thus 
$\partial_x\bar h= 
\partial_x \bar u/(1- \partial_z\bar u)$
and similarly for $\partial_y \bar h$.  For low-angle
grain-boundaries, $\partial_z \bar u\approx 0$ and the derivatives of $\bar h$ and
$\bar u$ are
equal.  This will not be the case for {\sl large} angle grain-boundaries.
With this in mind, we will add together
the effect of a plane of equally spaced screw dislocations with spacing $l_d$.

A single screw dislocation is described by $\bar u_1=[b/(2\pi)]\tan^{-1}(y/x)$ with
$b=kd$:
\eqn\etilts{\left[\partial_x \bar u_1,\partial_y \bar u_1,\partial_z \bar u_1\right]
= {b\over 2\pi}\left[{-y\over x^2 + y^2},{x\over x^2 + y^2},0\right].}
An array of screw dislocations gives:
\eqn\earray{\eqalign{
\partial_x \bar u_{\rm array} &= -{b\over 2\pi}\sum_{n=-\infty}^\infty 
{y-nl_d\over x^2 + (y-nl_d)^2}\cr
\partial_y \bar u_{\rm array} &= {b\over 2\pi}\sum_{n=-\infty}^\infty 
{x\over x^2 +(y-nl_d)^2}.\cr
}}
These sums can be performed explicitly via the Poisson summation formula:
\eqn\esoln{\eqalign{
\partial_x \bar u_{\rm array} &= -{b\over 2l_d}{\sin(2\pi y/l_d)\over
\cosh(2\pi x/l_d) - \cos(2\pi y/l_d)}\cr
\partial_y \bar u_{\rm array} &= {b\over 2l_d}{\sinh(2\pi x/l_d)\over
\cosh(2\pi x/l_d) - \cos(2\pi y/l_d)}.\cr}}
Note that the superposition of \etilts\ gives $\partial_z\bar u_{\rm array}=0$, and
thus $\bar u_{\rm array}$ is independent of $z$.
However, $\bar u_{\rm array}$
is the displacement due only to the screw dislocations -- the smectic
will relax via a smooth displacement field $\bar u_{\rm smooth}(x,y,z)$.  To find
$\bar u_{\rm smooth}$ we consider the boundary condition on the smectic: 
the strain must vanish at $x=\pm\infty$.
Writing $\bar u= \bar u_{\rm
array} + \bar u_{\rm smooth}$, the rotationally invariant strain is $u_{zz}=\partial_z \bar u - {1\over 2}
\left(\nabla
\bar u\right)^2$ and the boundary condition becomes:      
\eqn\epminfty{2\partial_z \bar u - 
\left(\partial_z\bar u\right)^2 = \left({b\over 2l_d}\right)^2}
or $\partial_z \bar u = 1\pm\sqrt{1-(b/2l_d)^2}$.  Since $\bar u$ should
vanish as 
$l_d\rightarrow\infty$ (or $b\rightarrow 0$), 
we have $\partial_z\bar u = 1-\sqrt{1-(b/2l_d)^2}$.  Comparing \escherkdif\ and
\esoln\ while translating from Eulerian to Lagrangian co\"ordinates, we find that Scherk's surface is an array
of screw dislocations, {\sl i.e.}, a TGB with:
\eqna\escherkrules{
$$\eqalignno{\tan(\alpha/2) &= \left(b/
2l_d\right)\left[1-\left(
b/ 2l_d\right)\right]^{-1/2},&\escherkrules a\cr
\sin(\alpha/2)&={\pi/l_d},&\escherkrules b\cr 
h(x,y) &= \bar h\big(x\cos(\alpha/2),y\big).&\escherkrules c\cr}$$}
Equation \escherkrules{a}\ implies that $\sin(\alpha/2) = b/(2l_d)$,
a standard result from the theory of twist-grain-boundaries, while
\escherkrules{b}\ shows that $b=2\pi=2d$, confirming that
there are strength $+2$ dislocations.  Finally, \escherkrules{c}\ displays
the only difference between Scherk's surface and the screw dislocations:
Scherk's surface is dilated along the twist axis by $\cos(\alpha/2)$.  We also note that
the Eulerian co\"ordinate for Scherk's surface is simply $u(x,y,z)=\bar
u(x\cos(\alpha/2),y,z)$.

We now consider the free energies of Scherk's surface and an LSD surface with
arbitrary Burgers vector $b=kd$. 
This free energy
has two contributions.  The first is the bending
energy.  In Lagrangian co\"ordinates it can be written as the product of the
free energy per surface, the density of surfaces per unit length and
the total length of the system.   We use the Helfrich-Cahn energy
for each surface, as has been utilized to study lyotropic monolayers 
\ref\WS{Z.-G.~Wang
and S.A.~Safran, J. Phys. (Paris) {\bf 51} (1990) 185.}. We find:
\eqn\ehc{F_{\rm b} = {\kappa L_z\over 2d\cos(\alpha/2)}\int
dxdy\sqrt{1+(\partial_xh)^2+(\partial_yh)^2}\, H^2\; ,} 
where $H$ is the mean curvature of the surface.  Note that $\kappa/d=K_1$ is the
three-dimensional splay modulus of the smectic.  
Scherk's surface is minimal and thus has $H=0$ -- which is precisely why
it proves to be a useful {\sl ansatz}.  The LSD surfaces
have a non-zero bending energy.  Using the standard non-linear expression
for $H$ \scherk , we find
\eqn\efb{F_{\rm b} = {\kappa L_z\Gamma^3\over 8d\cos(\alpha/2)}
\int d\tilde x d\tilde y \,
{\sinh^2(\tilde x)\sin^2(\tilde y)\left[\cosh(\tilde x) -
\cos(\tilde y)\right]^{-3/2}\over \left[\left(1+\Gamma\right)
\cosh(\tilde x)
- \left(1-\Gamma\right)\cos(\tilde y)\right]^{5/2}},  }
where $\Gamma=\tan^2(\alpha/2)$.
Inspection shows that \efb\ is a convergent integral, proportional
to the cross-sectional area of the sample, and that for
small angles $F_{\rm b}\sim b^4L_yL_z/(dl_d^5)$. 
Thus for large $l_d$, we see that the interaction energy between two
defects scales as $l_d^{-5}$.  This is different from the usual exponential
interactions that a linear theory with both director and displacement fields
\ref\lub{A.R.~Day, {\sl et al.}, Phys. Rev. A {\bf 27} (1983) 1461.}\ would predict.  
Moreover, for fixed angle $\alpha=2\sin^{-1}(b/2l_d)$,
larger values of $b$ are favored over smaller ones.
We note that in addition to the mean curvature term in \ehc ,
one could also consider a Gaussian curvature term of the form $\bar\kappa 
\int dS\, K$ where
$K$ is the Gaussian curvature.  
If the elastic constants do not depend on the co\"ordinates $(x,y)$, then
the integral of the Gaussian curvature is independent of the surface geometry.
The case of varying elastic constants has been considered elsewhere
\ref\ELTii{S.P.~Gido and E.L.~Thomas, Macromolecules {\bf 27} (1994) 849.}.
  
Since smectic and lamellar layers have a preferred spacing, there is
an energy cost for layer compression.
The compression energy $F_{\rm c}$
is most simply written in terms of the Eulerian co\"ordinates $u$:
\eqn\ecompress{F_{\rm c} = {B\over 2}\int d^3\!x\,\left[\partial_z u
- {1\over 2}\left(\nabla u\right)^2\right]^2,}
where the nonlinear terms assure complete rotational invariance.  
The compression energy does not vanish for either LSD surfaces
or for Scherk's surface.  
For Scherk's surface $F_{\rm c}$ 
is proportional to the cross-sectional area of
the sample in the $yz$-plane:
\eqn\efdiffii{
F_{\rm c} = {2BL_yL_z\sin^3(\alpha/2)\over \pi\cos(\alpha/2)}
\int_0^{\pi/2}\!\!\! d\theta \int_0^\infty \!\!\!d\gamma
\left[D\cos(2\theta) -{1\over 2}D^2\sin^2(\alpha/2)\sin^2(2\theta)\right]^2,}
where $D^{-1}=\left[\cosh(2\gamma)-\cos(2\theta)\right]\sim\left[\gamma^2
+\theta^2\right]$ for $\gamma,\theta\ll 1$. For an LSD surface
this compression energy $\bar F_{\rm c}$ is also proportional to the cross-sectional area:
\eqn\efdi{\bar F_{\rm c} = {BL_yL_z b^4\over \pi^2(2l_d)^3}\int_0^{\pi/2}   d\theta
\int_0^\infty d\gamma D^2\cos^2(2\theta).}
Note that the leading terms in the expansions of $F_{\rm c}$ and $\bar F_{\rm c}$ 
in $\alpha$ for $\alpha\ll 1$ are identical
when $b=2d$, {\sl i.e.}, for an LSD surface made of $+2$ dislocations.  Because
of the relative sign between the two terms in \efdiffii , it is difficult
to determine the relative magnitudes of $F_{\rm c}$ and $\bar F_{\rm c}$ for the
same twist angle $\alpha$ and $b=2d$. 
However, it is easy to see from \efdi\ 
that for small
angle, strength $+1$ dislocations lead to a {\sl lower} compression energy than
Scherk's surface.    

More importantly, in both cases (LSD and Scherk's surfaces), the integrals diverge
for small $\gamma$ and $\theta$.  This means that all of them must be cutoff
at some short distance $2\pi/\Lambda\sim d$ 
as has been observed by Kl\'eman \ref\KL{M.~Kl\'eman, 
{\sl Points, Lines, and Walls: in Liquid Crystals, Magnetic Systems, and
Various Ordered Media}, (Wiley, New York, 1983).}.  Inside the cutoff there is a defect core
in which the smectic order parameter must vanish.  The ``core energy'' that
we find here is actually a nonlinear {\sl elastic} energy, not to be confused with
the usual core energy which arises from the order parameter vanishing at
the defect and from relaxation of director modes.  We shall use the term ``core energy'' to refer to the elastic
core, keeping in mind that this is only part of the total core energy.
The integrals \efdiffii\ and \efdi\
diverge as $(\Lambda l_d)^2$.  Thus
the dominant contribution to the compression energy for $l_d\gg \Lambda^{-1}\sim d$ is 
\eqn\ecore{F_{\rm c} \sim {Bb^4L_z\over l_d^2}{L_y\over l_d} (\Lambda l_d)^2.}
This has a simple interpretation:  
since $L_y/l_d$ is the total number of dislocations, \ecore\ is 
the sum over individual dislocations of elastic
core energies $\epsilon L_z$ with energy per unit length
$\epsilon= Bb^4\Lambda^2$.
Unlike the bending energy, this energy strongly favors small Burgers vectors,.  
In addition, we may use \efdiffii\ or \efdi\ to calculate the leading
interaction energy between defects.  One finds that the interaction
energy is cutoff dependent through logarithmic terms.  
Therefore, on dimensional grounds the
interaction energy per unit length scales as $Bb^4l_d^{-2}\ln(\Lambda l_d)$.   
The cutoff dependence of the interaction is somewhat unusual: in linear
elasticity, defect--defect interactions only depend on the defect separation.
Because this system is non-linear, no such decomposition can be easily made.

Since both LSD and Scherk's structures necessarily have core regions, 
the energy of the core itself must
also be taken into account. 
A detailed analysis
of the core structure would be interesting and would allow for
explicit energy comparisons between the Scherk  and LSD structures. 
In such a calculation, the core size $2\pi/\Lambda$ itself
would be set so as to minimize the total energy.  This is similar to 
the analysis of defects in lyotropic lamellar systems \ref\KL{R.D.~Kamien and
T.C.~Lubensky, J. Phys. II France {\bf 7} (1997) 157.}.  
Note that there is
only one set of parallel core regions, despite the fact that there 
are two, perpendicular sets of helicoid-like structures in Scherk's surface \escherk. 
Once the core regions are present in one set of helicoids, the other set no longer 
consists of perfect helicoids -- they are interrupted by the cores of the first set.  
Thus, one
can unambiguously identify a single set of dislocations in the LSD and
Scherk structures.  This corroborates our earlier claim that
Scherk's surface contains a single set of true dislocations, while the
second set of (interrupted) helicoids appears only through the geometry of the other
defects.  Presumably, energy considerations will determine which set of helicoids
become dislocations.  At $\alpha=\pi/2$ the energy is degenerate.
However, at smaller $\alpha$ the preceding discussion suggests
that a system with a larger defect separation $l_d$ will have a lower energy.  This
choice of true topological defects
is consistent with the traditional construction of grain boundaries in
the TGB-A phase \TGB .
Finally, we note that Scherk's surface has a non-zero compression energy.
The true equilibrium twist grain boundary will adopt a geometry that is
an energetic compromise
between bending and compression deformations. As a result, it will have
a non-vanishing mean curvature, and it will have a structure that is identical
to neither Scherk's surface nor to any LSD surface discussed here. It would be interesting to consider
a variational {\sl ansatz} based on Scherk's surface with arbitrary, independent
dilations of $x$ and $y$.
  
We have shown that Scherk's surface is an anisotropic dilation of a periodic
surface constructed of a single set of strength $2$ screw dislocations.  Furthermore
since the lamellar ground state has a preferred layer spacing, layer
compression contributes to the free energy of the structure.  This
breaks the dual mapping between the helicoids. It thus
follows that Scherk's surface is a twist grain boundary composed of a {\sl single set} of parallel
screw dislocations and that the geometry of these defects creates a perpendicular
set of helicoidal structures in the surface.  We have also demonstrated that
Scherk's surface has
a higher energy than an LSD structure built of $+1$ dislocations for
small angles.  In the case of biphasic materials, the $+1$ dislocations
we consider here would be topologically forbidden -- 
they would become $+1/2$ dislocations.  In this case the energetics
would be a competition between Scherk's surface and the $+2$ LSD.  
In either case, a detailed analysis of the core structure would be required
to make an unambiguous prediction of the most stable structure at larger
angles, whether it is a distorted Scherk's surface,
a distorted LSD surface or
some interpolation between them.  
 
We acknowledge stimulating discussions with 
M.~Kl\'eman, B.~Pansu and E.L.~Thomas.
RDK thanks the Universit\'e Paris XI, Orsay, where some of this work
was done.  RDK was supported by
NSF CAREER Grant DMR97-32963, an award from Research Corporation and a gift
from L.J.~Bernstein.  TCL was supported by NSF
Grant DMR97-30405.

\listrefs
\listfigs

\bye